\documentclass{ws-procs9x6}            % default, citations in superscript
\begin{document}
\title{Light Meson Parton Distribution Functions from Basis Light-Front Quantization and QCD Evolution}

\author{Jiangshan Lan$^{a,b,c,*}$, Chandan Mondal$^{a,b}$, Shaoyang Jia$^{d}$,\\
 Xingbo Zhao$^{a,b}$ and James P. Vary$^{d}$\\
 (BLFQ Collaboration)}
\address{$^a$Institute of Modern Physics, Chinese Academy of Sciences, Lanzhou 730000, China\\
$^b$School of Nuclear Science and Technology, University of Chinese Academy of Sciences, Beijing 100049, China\\
$^c$Lanzhou University, Lanzhou 730000, China\\
$^d$Department of Physics and Astronomy, Iowa State University, Ames, IA 50011, USA\\
$^*$jiangshanlan@impcas.ac.cn}
%$^{\dag}$mondal@impcas.ac.cn\\
%$^{\ddag}$sjia@iastate.edu\\
%$^{\S}$xbzhao@impcas.ac.cn\\
%$^{\parallel}$jvary@iastate.edu}

\begin{abstract}
We investigate the parton distribution functions (PDFs) of the pion and kaon from the eigenstates of a light-front effective Hamiltonian in the constituent quark-antiquark representation suitable for low-momentum scale applications. By taking these scales as the only free parameters, the valence quark distribution functions of the pion, after QCD evolving, are consistent with the E615 experiment at Fermilab. In addition, the ratio of the up quark distribution in the kaon to that in the pion also agrees with the NA3 experimental result at CERN.
\end{abstract}

\keywords{BLFQ; LFWF; Meson; PDF.}

\bodymatter
\section{Introduction}\label{aba:sec1}
Basis light-front quantization (BLFQ) is a nonperturbative approach with advantage of both light-front dynamics and ab initio nuclear structure methods, which has been developed for solving many-body bound state problems in quantum field theories~\cite{Vary:2009gt,Wiecki:2014,Li:2015zda,Zhao:2014xaa}. It has been used to investigate a variety of systems and reactions~\cite{Li:2017mlw,Tang:2018myz,Chen:2018vdw,Li:2019kpr,Lan:2019img,Xu:2019xhk,Du:2019ips}. Recently, the BLFQ approach using a Hamiltonian that includes the color singlet Nambu--Jona-Lasinio (NJL) interaction to account for the chiral dynamics has been applied to light mesons \cite{Jia:2018ary,Lan:2019vui,Lan:2019rba}. In this work, we present results for the parton distribution functions (PDFs) of the light mesons.

	The probability of finding a valence quark with the longitudinal momentum fraction $x$ inside the meson ($|q\bar{q}\rangle$) is given by~\cite{Li:2017mlw}
	\begin{equation}\label{eq:PDF_valence}
\quad f(x)=\dfrac{1}{4\pi\,x(1-x)}\sum_{rs}\int \dfrac{d\overrightarrow{\kappa}^\perp}{(2\pi)^2}\,\psi^*_{rs}(x,\overrightarrow{\kappa}^\perp)\,\psi_{rs}(x,\overrightarrow{\kappa}^\perp),
	\end{equation}
where $\psi_{rs}(x,\overrightarrow{\kappa}^\perp)$ is the valence light-front wave function (LFWF) of the meson. Correspondingly, the valence antiquark PDF is given by ${f(1-x)}$. The momentum sum rule is
	\begin{equation}
	\int_{0}^{1}x\,f(x)\,dx+\int_{0}^{1}x\,f(1-x)\,dx=1,
	\end{equation}
which indicates that, at our model scale, the valence quark and antiquark carry the entire momentum of the meson. The normalization of our LFWFs ensures that valence quark PDFs are normalized. 

\section{Basis light-front quantization}\label{aba:sec2}
LFWFs encode the structure of bound states, in the BLFQ approach, one obtains the LFWF by solving the time-independent light-front Schr\"{o}dinger equation 
\begin{equation}
H_{\mathrm{eff}}\vert \Psi\rangle=M^2\vert \Psi\rangle,\label{eq:LF_Schrodinger}
\end{equation}
where $H_{\mathrm{eff}}$ is the effective Hamiltonian of the system with the eigenvalue of the mass squared $M^2$ . Generally, $\vert \Psi\rangle$ is expanded into all Fock sectors in the Hilbert space. In the valence Fock sector, the effective Hamiltonian for the light mesons with non-singlet flavor wave functions is given by~\cite{Jia:2018ary}
\begin{align}
H_\mathrm{eff} = \frac{\vec k^2_\perp + m_q^2}{x} + \frac{\vec k^2_\perp+m_{\bar q}^2}{1-x}
+ \kappa^4 \vec \zeta_\perp^2- \frac{\kappa^4~\partial_x\big( x(1-x) \partial_x \big)}{(m_q+m_{\bar q})^2} +H^{\rm eff}_{\rm NJL},\label{eqn:Heff}
\end{align}
where $m_q$ ($m_{\bar q}$) is the mass of the quark (antiquark), and $\kappa$ is the strength of the confinement. $\vec \zeta_\perp \equiv \sqrt{x(1-x)} \vec r_\perp$ is the holographic variable~\cite{Brodsky:2014yha}, with $\overrightarrow{k}_\perp$ being the conjugate variable of $\overrightarrow{r}_\perp$. The $x$-derivative is defined as $\partial_x f(x, \vec\zeta_\perp) = \partial f(x, \vec \zeta_\perp)/\partial x|_{\vec\zeta}$. 
The first two terms in Eq.~\eqref{eqn:Heff} are the light-front kinetic energy for the quark and the antiquark. The third and the fourth terms are the confining potentials in the transverse direction based on the light-front holographic QCD (LFHQCD)~\cite{Brodsky:2014yha} and a longitudinal confining potential~\cite{Li:2015zda} that reproduce 3D confinement in the nonrelativistic limit. Additionally, the $H_{\mathrm{NJL}}^{\mathrm{eff}}$ is the color-singlet NJL interaction to account for the chiral dynamics~\cite{Klimt:1989pm}.

\section{Results}\label{aba:sec3}
\begin{figure}
\begin{center}
\includegraphics[width=2.2in]{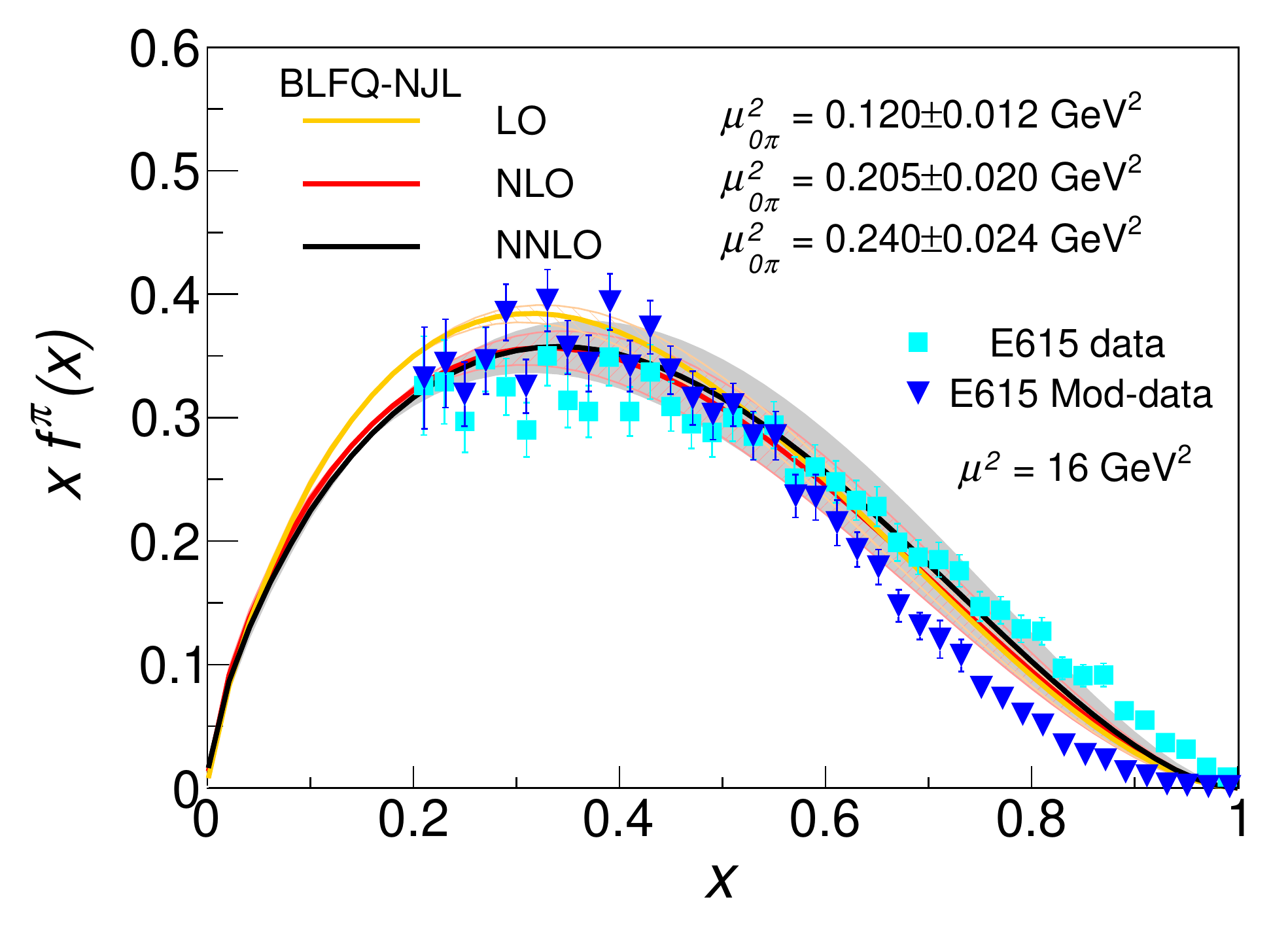}
\includegraphics[width=2.2in]{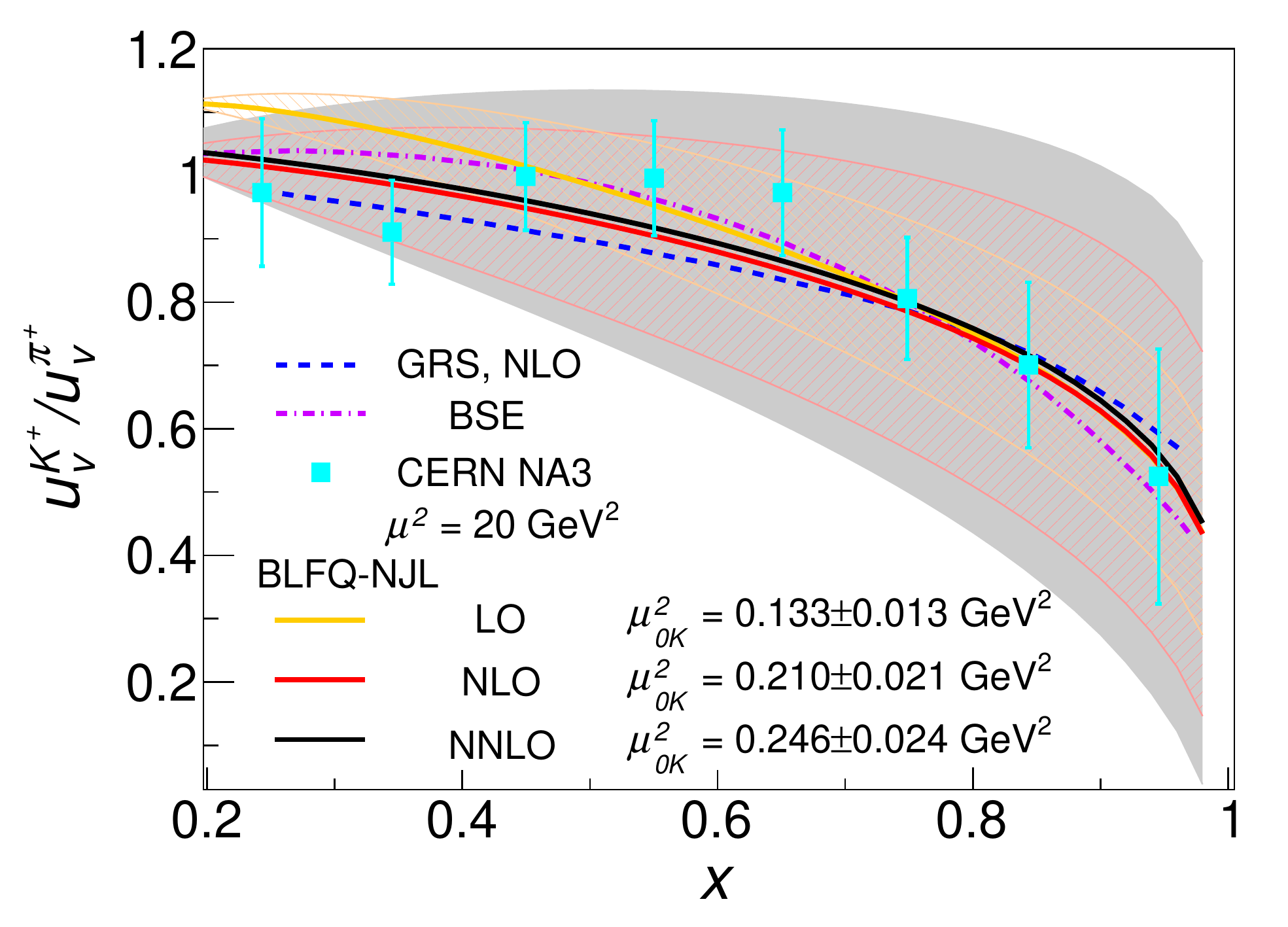}
\caption{(left) $xf^\pi(x)$ as a function of $x$ for the pion. (right) The ratio of the u quark PDFs in the kaon to that in the pion.}
\label{fig1}
	\end{center}
\end{figure}

We evolve our initial PDFs\cite{Lan:2019vui} to the relevant experimental scales ${\mu^2=16~\mathrm{GeV}^2}$ and ${\mu^2=20~\mathrm{GeV}^2}$ for the pion and the kaon by the DGLAP equations~\cite{Dokshitzer:1977sg,Gribov:1972ri,Altarelli:1977zs}, respectively. Here, we adopt the higher order perturbative parton evolution toolkit to solve the DGLAP equations numerically~\cite{Salam:2008qg}. We find that the initial scales increase when we progress from the leading order (LO) to the next-to-next-to-leading order (NNLO). Meanwhile, the evolved PDFs fit better to the experimental result demonstrated by smaller values of $\chi^2$ per degree of freedom (d.o.f.) at higher orders, as shown in Table \ref{table1} as well as Figure \ref{fig1}. %Since the results from the higher order DGLAP equation appear more reliable due to higher initial scales, only the results for the PDFs at NNLO are presented in this paper. 

\begin{table}
\tbl{Initial scales and the $\chi^2$/(d.o.f.) at the first three orders of the DGLAP equation.}
{\begin{tabular}{@{}ccccc@{}}\toprule
Order &$\mu^2_{0,\pi}~ [\mathrm{GeV}^2]$& $\mu^2_{0,K}~ [\mathrm{GeV}^2]$ & E615\cite{Conway:1989fs} $\frac{\chi^2}{(d.o.f.)}$ & NA3\cite{Badier:1983mj} $\frac{\chi^2}{(d.o.f.)}$ \\ 
\colrule
LO & $0.120\pm 0.012$ & $0.133\pm 0.013$ & $6.71$ & $0.88$ \\ 
NLO & $0.205\pm 0.020$ & $0.210\pm 0.021$ & $4.67$ &$0.56$ \\ 
NNLO & $0.240\pm 0.024$ & $0.246\pm 0.024$ & $3.64$ & $0.50$ \\ 
\botrule
\end{tabular}
}
\label{table1}
\end{table}

%\begin{table}
%\tbl{The contributions to the first moment of the pion.}% at 4 GeV$^2$
%{\begin{tabular}{@{}cccc@{}}\toprule
%$\langle x \rangle$ (4 GeV$^2$) & Valence & Gluon & Sea \\
%\colrule
%BLFQ-NJL & $0.489$  &  $0.398$  &  $0.113$  \\
%BSE~\cite{Ding:2019lwe} &    $0.48(3)$  &  $0.41(2)$  &  $0.11(2)$ \\
%\botrule
%\end{tabular}
%}
%\label{table2}
%\end{table}
In Table \ref{table1}, the $\chi^2$ is defined as the sum of the squared-difference of our results from the results of the E615 experiment at FNAL \cite{Conway:1989fs} and from the NA3 experiment at CERN \cite{Badier:1983mj}, both at the respective experimental scales.
In Figure \ref{fig1} (left panel), we show our results in each order from LO to NNLO compared with the E615 experimental result \cite{Conway:1989fs} and the modified result of the E615 experiment \cite{Chen:2016sno} for the pion. In Figure 1 (right panel) we compare our results for the ratio of the kaon to the pion up quark distributions with the NA3 experimental result \cite{Badier:1983mj} and the NLO Gl\"{u}ck-Reya-Stratmann (GRS) model \citep{Gluck:1997ww} as well as the prediction from the  Bethe-Salpeter equation (BSE) \cite{Nguyen:2011jy} for the Kaon. The bands are due to a 10\% uncertainty for initial scales. At large $x$, the behavior of our pion valence PDF is $(1-x)^{1.44}$. Furthermore, the various valence, gluon, and sea contributions to the first moment of the pion at 4 GeV$^2$ are \{0.489, 0.398, 0.113\}, respectively, which agree with the results from BSE model~\cite{Ding:2019lwe} \{0.48(3), 0.41(2), 0.11(2)\}. We further find our results agree well with other results \cite{Conway:1989fs,Badier:1983mj,Chen:2016sno,Gluck:1997ww,Nguyen:2011jy,Ding:2019lwe}.

\section{Conclusions}\label{aba:sec4}
The PDFs of the light mesons have been presented using the BLFQ approach and QCD evolution. In our model, the initial scales of PDFs up to NNLO are ${\mu_{0\pi}^2=0.240\pm0.024~\rm{GeV}^2}$ for the pion and ${\mu_{0K}^2=0.246\pm0.024~\rm{GeV}^2}$ for the kaon. The result of the pion valence PDF has produced good agreement with the E615 experimental result \cite{Conway:1989fs} and the modified result of the E615 experiment \cite{Chen:2016sno}. In addition, the contributions to pion first moment at 4 GeV$^2$ is consistent with the results from the BSE model~\cite{Ding:2019lwe}. We also note that the ratio of the up quark PDFs in the kaon to that in the pion is in good agreement with the CERN-NA3 experimental result~\cite{Badier:1983mj}, the result from GRS model \citep{Gluck:1997ww} as well as the prediction from the BSE \cite{Nguyen:2011jy}. 
\section*{Acknowledgments}
CM is supported by the National Natural Science Foundation of China (NSFC) under the Grant No. 11850410436. XZ is supported by new faculty startup funding by the Institute of Modern Physics, Chinese Academy of Sciences and by Key Research Program of Frontier Sciences, CAS, Grant No ZDBS-LY-7020.  SJ and JPV are supported in part by the Department of Energy under Grants No. DE-FG02-87ER40371 and No. DESC00018223 (SciDAC-4/NUCLEI). A portion of the computational resources were provided by the National Energy Research Scientific Computing Center (NERSC), which is supported by the Office of Science of the U.S. Department of Energy under Contract No. DE-AC02-05CH11231.

\bibliographystyle{ws-procs9x6} % for numbered citation & references

\end{document}